\documentclass[fleqn,twoside]{article}
\usepackage{espcrc2}

\usepackage{graphicx}
\usepackage[figuresright]{rotating}

\newcommand{\AmS}{{\protect\the\textfont2
  A\kern-.1667em\lower.5ex\hbox{M}\kern-.125emS}}
  
  \usepackage{amsmath}
\usepackage{amssymb}
  
\def\be{\begin{equation}}
\def\ee{\end{equation}}
\def\ba{\begin{eqnarray}}
\def\ea{\end{eqnarray}}

\def\f{\frac}

\def\ie{{\frenchspacing\it i.e.}}

\title{Constraining modified growth patterns with tomographic surveys}

\author{Alessandra Silvestri\address{Physics Department, Massachusetts Institute of Technology, 
        Cambridge, MA, 02139, USA}
        \thanks{collaborators on the work presented: GongBo Zhao, Levon Pogosian and Joel Zylberberg}}
       
\begin{document}

\begin{abstract}
Viable models of modified gravity designed to produce cosmic acceleration
at the current epoch, closely mimic the $\Lambda$CDM model at the level of background
cosmology. However, this degeneracy is generically broken at the level of linear perturbations,
where the modifications induce a peculiar scale-dependent pattern. A similar pattern is expected
in models of coupled dark energy. I present the main results published in Pogosian and Silvestri~\cite{Pogosian:2007sw}, on the growth of structure in $f(R)$ theories of gravity, and in Zhao {\it et al.}~\cite{Zhao:2008bn}, on the potential of upcoming and future tomographic surveys to detect departures  from the growth of cosmic structure expected within General Relativity with a  cosmological constant.
\vspace{1pc}
\end{abstract}

% typeset front matter (including abstract)
\maketitle

\section{Introduction}
\label{Intro}
Observations strongly favor a universe that has recently entered a phase of accelerated expansion~\cite{Riess:1998cb,perlmutter}. This poses a puzzle for Modern Cosmology since standard \emph{General Relativity} (GR), applied to a universe which contains only radiation and dust, has difficulties fitting the data. The cosmological constant $\Lambda$, a dark energy (DE) fluid (i.e. a yet unknown component with a negative equation of state) as well as modifications of GR on large scales are the different proposals  to address the phenomenon of cosmic acceleration.

It has become increasingly evident that tests based solely on the expansion history cannot place significant bounds on many models that address cosmic acceleration. Indeed, although current geometrical probes place tight constraints on the expansion history, models of modified gravity in general have enough freedom to reproduce any desired expansion history. In other words, at the level of background cosmology  there is a degeneracy among different models of cosmic acceleration~\cite{Pogosian:2007sw,Dvali:2007kt,Hu:2007nk,Appleby:2007vb,Brax:2008hh,Capozziello:2008fn}. 
However, this degeneracy is typically broken at the level of cosmological structure formation; indeed, models of modified gravity that closely mimic the cosmological constant at the background level can still give significantly different predictions for the growth of structure.  

An instructive example is offered by $f(R)$ theories of gravity. These models are degenerate with $\Lambda$CDM (i.e. GR with cold dark matter and a cosmological constant) at the level of background cosmology. However, they predict a significantly different  dynamics of perturbations, with a peculiar scale-dependent pattern for the growth of structure. 
Studying in detail the formation of structure in this class of models allows one to gain insight on the growth of perturbations in generalized models in which a scalar field, (being it the dark energy field or the additional degree of freedom introduced by modifications of gravity), is coupled to dark matter. 

From this analysis it becomes clear that the large scale structure of the universe offers a promising testing ground for GR. It is therefore important to explore to what extent one can detect departures from GR in the growth of structure with present and upcoming cosmological data. In this regard, I review the results of Zhao {\it et al.}~\cite{Zhao:2008bn}, where we have used parametric forms to quantify the potential time- and scale-dependent modifications of the growth of structure. Since the effects of the modifications of gravity on the growth of structure are typically equivalent to, or can be mimicked by, a more exotic dark energy field, (i.e. one that couples to dark matter and/or carries anisotropic stress), we have focused on the distinction between $\Lambda$CDM and any alternative.

%%%%%%%%%%%%%%%%%%%%%%%%%%%%%%%%%%%%%%%%%%%%%%%%%%%
\section{$f(R)$ theories of gravity}\label{f_R}
In 1979 Starobinsky showed that a de Sitter phase in the early universe can be achieved by adding to the Einstein action a function $f(R)\propto R^2$~\cite{Starobinsky:1979ty,Starobinsky:1980te}. More recently, $f(R)$ theories have been revisited in the context of cosmic acceleration, starting with~\cite{Capozziello:2003tk,Carroll:2003wy} and followed by many others. The underlying idea is to add functions of the Ricci scalar which become important at late times, for small values of the curvature. Let us consider the action for a generic $f(R)$ model
\ba
S&=&\frac{1}{2\kappa^{2}}\int d^4 x\sqrt{-g}\, \left[R+f(R)\right] \nonumber\\
&&+ \int d^4 x\sqrt{-g}\, {\cal L}_{\rm m}[\chi_i,g_{\mu\nu}] \,,
\label{jordanaction}
\ea
where $\chi_i$ represents the matter fields, which are minimally coupled and, therefore, fall along geodesics of the metric $g_{\mu\nu}$. The field equations obtained varying the action~(\ref{jordanaction}) with respect to $g_{\mu\nu}$, and the energy-momentum conservation equations are, respectively,
 \ba \label{jordaneom1}
&&\left(1+f_R \right)R_{\mu\nu} - \frac{1}{2} g_{\mu\nu} \left(R+f\right)\nonumber\\
&& + \left(g_{\mu\nu}\Box 
-\nabla_\mu\nabla_\nu\right) f_R =\kappa^{2}T_{\mu\nu} \ ,\\
\nonumber\\
\label{jordaneom2}&&\nabla_{\mu}T^{\mu\nu}=0\,,
\ea
where $f_R\equiv df/dR$. The Einstein equations~(\ref{jordaneom1}) are now fourth order in the scale factor. The additional dynamics introduced by the modifications of gravity can be seen explicitly by taking the trace of Einstein equations, to get
\be
\Box{f}_R={1\over 3}\left(R+2f-Rf_R \right)-{\kappa^2 \over 3}(\rho-3P) 
\equiv {\partial V_{\rm{eff}} \over \partial f_R} \ ,
\ee
which is a second order equation for $f_R$, with a canonical kinetic term and an effective  potential $V_{\rm{eff}}(f_R)$. 
Therefore the modifications of the Einstein action introduce an  additional scalar degree of freedom and the role of the scalar field is played by $f_R$, dubbed {\it the scalaron}~\cite{Starobinsky:1980te}. One can associate the following Compton wavelength to the scalaron
\be\label{Compton_f(R)}
\lambda_c\approx 2\pi\sqrt{\frac{3f_{RR}}{1+f_R}}
\ee
The scalaron adds extra dynamics to the theory and mediates a fifth-force between matter fields. Both these features can be problematic for the viability of the $f(R)$ model. Consequently, special care needs to be taken in order to ensure  a stable high-curvature regime, to have a positive effective Newton's constant and to satisfy local tests of gravity~\cite{Hu:2007nk,Dolgov:2003px,Amendola:2006kh,Navarro:2006mw,Chiba:2006jp,Amendola:2007nt,Starobinsky:2007hu}. After all the necessary conditions are imposed, viable $f(R)$ models closely mimic $\Lambda$CDM at the level of expansion history, with undistinguishable departures from $w_{\rm eff}=-1$. Nevertheless the predictions for the growth of structure can differ significantly from those of $\Lambda$CDM, as we will see in the following subsection.

\subsection{Dynamics of Linear Perturbations}
Let us concentrate on linear scalar perturbations to a flat Friedmann-Robertson-Walker universe, in Newtonian gauge. In particular, since we are interested in the growth of structure, let us focus on cold dark matter and consider linear sub-horizon scales. In Fourier space, the equations governing the evolution of perturbations are
\ba\label{f(R)_pert_eqs1}
&&\delta''+\left(1+\f{H'}{H}\right)\delta'+\f{k^2}{a^2H^2}\Psi=0\\
\label{f(R)_pert_eqs2}
&&\f{k^2}{a^2}\Psi=-\f{a^2\rho}{2M_P^2}\f{1}{1+f_R}\f{1+4\f{k^2}{a^2}\f{f_{RR}}{1+f_R}}{1+3\f{k^2}{a^2}\f{f_{RR}}{1+f_R}}\delta\\
\label{f(R)_pert_eqs3}
&&\f{\Phi}{\Psi}=\f{1+2\f{k^2}{a^2}\f{f_{RR}}{1+f_R}}{1+4\f{k^2}{a^2}\f{f_{RR}}{1+f_R}}
\ea
where a prime indicates derivation w.r.t. $\ln a$, $\delta\equiv \delta\rho/\rho$ is the density contrast of dark matter and $\Psi$ and $\Phi$ are, respectively, the perturbations to the time-time and space diagonal metric component.

\begin{figure}[t]
\includegraphics[width=1\columnwidth]{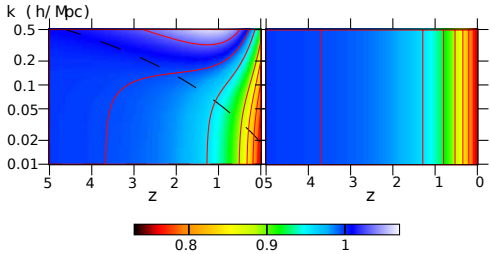}
\caption{The evolution of the growth factor for the CDM $[\Delta(k,a)/a]/[\Delta(k,a_i)/a_i]$ as a function of redshift and scale. The left panel corresponds to an $f(R)$ model with $w_{\rm{eff}}=-1$ and $\lambda_c^0\approx10^{2}\,{\rm Mpc}$, the right one to $\Lambda$CDM. The dashed line crossing diagonally on the left plot corresponds to the Compton scale (\ref{Compton_f(R)}).} \label{fig:mu_gamma}
\end{figure}

Comparing eq.~(\ref{f(R)_pert_eqs2}) with~(\ref{Compton_f(R)}), it is clear that the modifications have introduced a Yukawa-type correction to the Newtonian potential, with characteristic lengthscale $\lambda_c$ (\ref{Compton_f(R)}): the scalaron mediates a fifth force. Therefore, the Compton wavelength of the  scalaron introduces a scale which separates two regimes of sub-horizon gravitational dynamics during which gravity behaves differently, as can be noticed in Fig.~\ref{fig:mu_gamma}. On scales $\lambda\gg\lambda_c$, the scalaron is massive and the fifth force is exponentially suppressed, thus deviations from GR are negligible. However, on scales inside the Compton radius, the scalaron is light and deviations are significant. The relations between $\Phi$ and $\Psi$, and the relation between them and the matter density contrast, will be different below the Compton scale and that affects the growth rate of structures. In particular, on scales below $\lambda_c$, the modifications introduce a slip between these potentials, leading to $\Psi\simeq 2\Phi$. For modes that cross below $\lambda_c$ during matter domination the effect of modifications is maximized as the sum of the potentials grows in the absence of background acceleration. When, however, the background starts accelerating, the potentials begin to decay but at a lesser rate than in the $\Lambda$CDM model. A characteristic signature of $f(R)$ theories would be a non-zero Integrated Sachs-Wolfe effect (ISW) during the matter era. 

Overall, we observe a scale-dependent growth pattern. The modifications introduced by $f(R)$ models are analogous to those introduced by more general scalar-tensor theories and models of dark energy coupled to dark matter. The dynamics of perturbations is richer, with different observables described by different functions, not by a single growth factor. Therefore combining different measurements, such as galaxy counts, weak lensing and their cross-correlation with the ISW effect, we can test gravity on cosmological scales. But what is the power, of upcoming and future tomographic surveys to constrain modifications from $\Lambda$CDM? In~\cite{Zhao:2008bn}, the authors addressed this question employing a parametrization to describe the modifications, i.e. the rescaling of the Newton's constant and the gravitational slip. In what follows I review the main results of~\cite{Zhao:2008bn}.

%%%%%%%%%%%%%%%%%%%%%%%%%%%%%%%%%%%%%%%%%%%%%%%%%%%
\section{Searching for modified growth patterns}\label{MGR_obs}
The linearized Einstein equations provide two independent equations relating the metric potentials and matter perturbations: the {\it Poisson} and {\it anisotropy} equations. Just like in $f(R)$, eqs.~(\ref{f(R)_pert_eqs2}) and~(\ref{f(R)_pert_eqs3}), in models of modified gravity and more exotic models of dark energy, the relation between the two Newtonian potentials, and between the potentials and matter perturbations, can differ from the $\Lambda$CDM case~\cite{Zhang:2005vt,Bertschinger:2006aw,Song:2006ej}. We parametrize the changes to the Poisson and the anisotropy equations as follows:
\ba\label{parametrization-Poisson}
&&k^2\Psi=-\f{a^2}{2M_P^2}\mu(a,k)\rho\Delta\\
\label{parametrization-anisotropy}
&&\f{\Phi}{\Psi}=\gamma(a,k) \ ,
\ea
where $\rho\Delta\equiv\rho\delta+3\f{aH}{k}(\rho+P)v$ is the comoving density perturbation, $\mu(a,k)$ and $\gamma(a,k)$ are two time- and scale-dependent functions encoding the modifications of gravity and/or the contribution of an exotic dark energy fluid. In $\Lambda$CDM $\mu=1=\gamma$.

To arrive at a suitable parametrization of the functions $\mu(a,k)$
and $\gamma(a,k)$, we note that models of modified gravity typically
introduce a transition scale which separates regimes where gravity
behaves differently. For instance in $f(R)$ theories the functions $\mu$ and $\gamma$ are equal to unity at early
times and on large scales ($\ie$ on scales that are larger than the
Compton scale, $\lambda_c$, of the model), and they transition to a
modified value on smaller scales and late times. This
time- and scale-dependent transition can be mimicked via the following functions
\ba\label{par_G}
&&\mu(a,k)=\f{1+\beta_1\lambda_1^2\,k^2a^s}{1+\lambda_1^2\,k^2a^s}\\
\label{par_gamma}
&&\gamma(a,k)=\f{1+\beta_2\lambda_2^2\,k^2a^s}{1+\lambda_2^2\,k^2a^s} \ ,
\ea
where the parameters $\lambda^2_i$ have dimensions of length squared, while the $\beta_i$ represent dimensionless couplings and $s>0$. Expressions~(\ref{par_G}) and~(\ref{par_gamma}) coincide with the scale-dependent parametrization introduced in~\cite{Bertschinger:2008zb}. It is easy to see that this parametrization corresponds to the addition of a Yukawa-type interaction between dark matter particles, with an effective coupling $(\beta_1-1)$ and a range $\lambda_1$. In the case of $f(R)$ indeed we have seen that the modifications introduce a fifth-force mediated by the scalaron.

In order to determine how well upcoming and future surveys can constrain the Modified Growth (MG) parameters \{$s,\beta_1,\beta_2,\lambda_1,\lambda_2$\}  we employ the standard Fisher matrix technique~\cite{Fisher}. 

As our observables, we use all possible two-point correlation functions (both auto- and cross-correlations)
between the \emph{Galaxy Counts} (GC), \emph{Weak Lensing shear} (WL), and \emph{Cosmic Microwave Background} (CMB) temperature anisotropy, across multiple redshift bins, in addition to the CMB E-mode polarization autocorrelation, and the CMB E-mode and temperature cross-correlation. 

In order to compute the theoretical expectations for these observables, we assume fiducial values for the parameters and evolve the perturbations using MGCAMB\footnote{http://www.sfu.ca/$\sim$gza5/Site/MGCAMB.html}\cite{Zhao:2008bn}, (a modified version of the publicly available CAMB~\cite{camb,Lewis:1999bs}). For the fiducial values of the parameters we consider several models based on $f(R)$ and Chameleon theories. All $f(R)$ models correspond to a fixed coupling, $\beta_1=4/3\,(=8/3\,\beta_2)$ and to $s\sim 4$. The only free parameter is the lengthscale $\lambda_2=\sqrt{\beta_1}\lambda_1$. Here I report results for the specific choice of $\lambda_2^2=10^4\,{\rm Mpc}^2$.
In addition to the effects of modified gravity, the evolution of all cosmological perturbations depends on the background expansion. We restrict ourselves to background histories consistent with the flat $\Lambda$CDM model, since that is currently the best fit to available data and since popular models of modified gravity give an expansion history which is effectively the same as in $\Lambda$CDM; the main differences between models indeed arise from the evolution of cosmic structure. We do not, however, fix the values of the cosmological parameters $\Omega_m h^2$, $\Omega_b h^2$ and $h$, nor the spectral index or the optical depth. We vary them along with the modified growth parameters, using the WMAP 5-year data best fit~\cite{Dunkley:2008ie} for their fiducial values.

Finally, we assume CMB T and E data from the Planck satellite~\cite{Planck}, the galaxy
catalogues and WL data by the \emph{Dark Energy Survey} (DES)~\cite{DES} and \emph{Large Synoptic Survey Telescope} (LSST)~\cite{LSST}, complemented by a futuristic SNe data set provided by the ongoing \emph{Canada-France-Hawaii Telescope's}  \emph{Supernovae Legacy Survey}~\cite{SNLS,Astier:2005qq}, the \emph{Nearby Supernovae Factory}~\cite{NSNF,aldering02}, and a future \emph{Joined Dark Energy Mission} space mission, such as the \emph{Supernovae/Acceleration Probe}~\cite{SNAP}. We describe the specifics of the experiments in~\cite{Zhao:2008bn}.

%%%%%%%%%%%%%%%%%%%%%%%%%%%%%%%%%%%%%%%%%%%%%%%%%%%
\subsection{Results}
\label{results}
\begin{figure}[tb]
\includegraphics[width=1.\columnwidth]{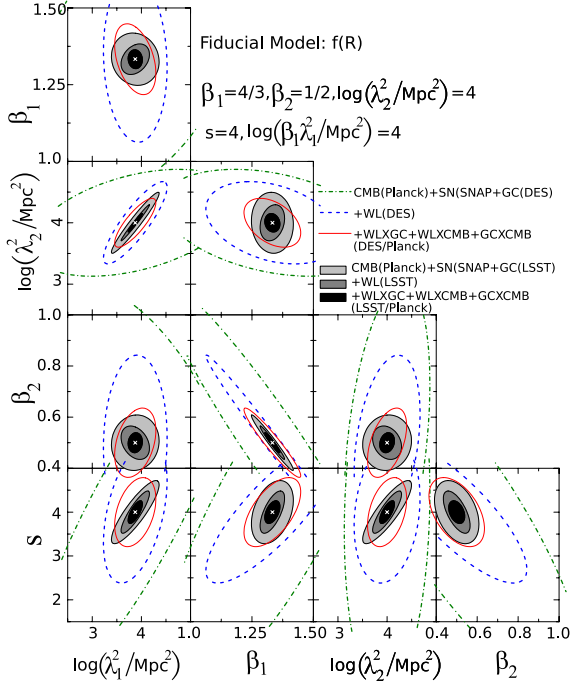}
\caption{The $68$\% confidence contours for the five MG parameters for the fiducial $f(R)$ model with $\lambda_2^2=10^4\,{\rm Mpc}^2$, as constrained by several different combinations of correlation functions from various experiments (shown in the legend). Each of the ellipses is plotted after marginalizing over all other parameters that were varied.} \label{fig:model2}
\end{figure}
We evaluate the Fisher
errors on the MG parameters, as well as the usual set of standard cosmological parameters. Fig.~\ref{results} shows the $68\%$ C.L. contours for the fiducial case of an $f(R)$ model with Compton scale $\lambda_2^2=10^4\,{\rm Mpc}^2$. This is just one representative case among  the several fiducial models considered in our analysis~\cite{Zhao:2008bn}. 
In general, we find that the five parameters are correlated and the degeneracy between $\lambda_1$ and $\lambda_2$ is
strongly positive, while $\beta_1$ and $\beta_2$ have a strong negative correlation. This is expected since one can enhance the growth by either raising $\beta_1$, which increases the effective
Newton's constant, or lowering $\beta_2$, which enhances the relative strength
of the Newtonian potential $\Psi$ which drives the clustering of matter. This also explains the degeneracy between the two length scale parameters $\lambda_1$ and $\lambda_2$.

Comparing the constraints on the MG parameters from different datasets, we see that the constraining powers
of WL, GC and WL$\times$GC are comparable and that they provide much more stringent constraints than WL$\times$CMB and GC$\times$CMB. This is, in part, because there are many more different
correlations between WL and GC than there are cross-correlations with CMB. Also, correlations of GC and WL with CMB suffer from a larger statistical uncertainty, since only the ISW part of the total CMB
anisotropy is correlated with large scale structure. 

Comparing the constraints from DES with LSST, we find that,
typically, LSST can improve constraints on individual MG parameters by a factor of $3$ or better.
With all tomographic data combined (no priors used), DES give relative errors (i.e. 
the $1\sigma$ marginalized
errors divided by the corresponding fiducial values) on the MG parameters of order $20$\%-$40$\%, while
LSS can go below $10$\% levels. 
%%%%%%%%%%%%%%%%%%%%%%%%%%%%%%%%%%%%%%%%%%%%%%%%%%%
\section{Conclusions}\label{summary}
Models of modified gravity and more exotic dark energy typically induce a scale-dependent growth pattern and a time- and scale-dependent slip between the gravitational potentials. Studying in detail the class of $f(R)$ theories, we have learned how the modifications come into play and which peculiar signatures they imprint on the LSS and CMB. In general, the dynamics of perturbations is richer, breaking the degeneracy that characterizes cosmic acceleration models at the background level.  Weak lensing, galaxy count, the ISW effect and their cross-correlations offer a powerful testing ground for GR on large scales. It is therefore important to determine to which extent current and future data can detect departures from GR in the dynamics of perturbations. 

In~\cite{Zhao:2008bn} we have used a five-parameters description for the rescaling of the Newton constant, $\mu(a,k)$, and for the ratio of the metric potentials, $\gamma(a,k)$, to investigate the power of upcoming and future tomographic surveys to constrain departures from GR. From the point of view of scalar-tensor theories, e.g. $f(R)$, these parameters are  related to the coupling in the dark sector and to the characteristic mass scale of the model.  We have then studied in detail the constraints on the five-parameters based on several fiducial models, expected from Weak Lensing, Galaxy Counts, CMB and their cross-correlation spectra as seen by Planck, DES and
LSST (additionally using Planck and SNAP to constrain the standard set of cosmological parameters). We have found that DES can provide $20$\%-$40$\% level constraints on individual parameters, and that LSST can improve on that, constraining the MG parameters to better than $10$\% level. Overall, with DES, and especially with LSST, one will be able to
significantly reduce the volume of the allowed parameter space in scalar-tensor type models.

From this analysis we have learned that upcoming and future surveys can place non-trivial bounds on modifications of the evolution of perturbations, even in the most conservative case, i.e. considering only linear scales as we did. These results are model-dependent, but they motivated us to pursue model-independent methods such as the Principal Component Analysis~\cite{tegmark-pca}; we recently reported the first results in~\cite{Zhao:2009fn}.

{\bf Acknowledgments} I would like to thank the organizers for a very enjoyable workshop and the Galileo Galilei Institute for hospitality.
%%%%%%%%%%%%%%%%%%%%%%%%%%%%%%%%%%%%%%%%%


\begin{thebibliography}{99}
%
  
  %\cite{Pogosian:2007sw}
\bibitem{Pogosian:2007sw}
  L.~Pogosian and A.~Silvestri,
  %``The pattern of growth in viable f(R) cosmologies,''
  Phys.\ Rev.\  D {\bf 77}, 023503 (2008)
  [arXiv:0709.0296 [astro-ph]].
  %%CITATION = PHRVA,D77,023503;%%
  
 
%\cite{Zhao:2008bn}
\bibitem{Zhao:2008bn}
  G.~B.~Zhao, L.~Pogosian, A.~Silvestri and J.~Zylberberg,
  %``Searching for modified growth patterns with tomographic surveys,''
  arXiv:0809.3791 [astro-ph].
  %%CITATION = ARXIV:0809.3791;% 
  
%\cite{Riess:1998cb}
\bibitem{Riess:1998cb}
  A.~G.~Riess {\it et al.}  [Supernova Search Team Collaboration],
  %``Observational Evidence from Supernovae for an Accelerating Universe and a
  %Cosmological Constant,''
  Astron.\ J.\  {\bf 116}, 1009 (1998)
  [arXiv:astro-ph/9805201].
  %%CITATION = ANJOA,116,1009;%%

\bibitem{perlmutter} S.~Perlmutter {\it et al}, Astrophys.~J.~{\bf 517}, 565 (1999).

\bibitem{Will} C.~M.~Will, {\it Theory and experiment in gravitational physics}, revised edition, Cambridge University Press (1993).

%\cite{Will:2005va}
\bibitem{Will:2005va}
  C.~M.~Will,
  %``The confrontation between general relativity and experiment,''
  Living Rev.\ Rel.\  {\bf 9}, 3 (2005)
  [arXiv:gr-qc/0510072].
  %%CITATION = 00222,9,3;%%


%\cite{Dvali:2007kt}
\bibitem{Dvali:2007kt}
  G.~Dvali, S.~Hofmann and J.~Khoury,
  %``Degravitation of the cosmological constant and graviton width,''
  Phys.\ Rev.\  D {\bf 76}, 084006 (2007)
  [arXiv:hep-th/0703027].
  %%CITATION = PHRVA,D76,084006;%%

%\cite{Hu:2007nk}
\bibitem{Hu:2007nk}
  W.~Hu and I.~Sawicki,
  %``Models of f(R) Cosmic Acceleration that Evade Solar-System Tests,''
  Phys.\ Rev.\  D {\bf 76}, 064004 (2007)
  [arXiv:0705.1158 [astro-ph]].
  %%CITATION = PHRVA,D76,064004;%%

%\cite{Appleby:2007vb}
\bibitem{Appleby:2007vb}
  S.~A.~Appleby and R.~A.~Battye,
  %``Do consistent $F(R)$ models mimic General Relativity plus $\Lambda$?,''
  Phys.\ Lett.\  B {\bf 654}, 7 (2007)
  [arXiv:0705.3199 [astro-ph]].
  %%CITATION = PHLTA,B654,7;%%

%\cite{Brax:2008hh}
\bibitem{Brax:2008hh}
  P.~Brax, C.~van de Bruck, A.~C.~Davis and D.~J.~Shaw,
  %``f(R) Gravity and Chameleon Theories,''
  arXiv:0806.3415 [astro-ph].
  %%CITATION = ARXIV:0806.3415;%%

   %\cite{Capozziello:2008fn}
\bibitem{Capozziello:2008fn}
  S.~Capozziello, M.~De Laurentis, S.~Nojiri and S.~D.~Odintsov,
  %``$f(R)$ gravity constrained by PPN parameters and stochastic background of
  %gravitational waves,''
  arXiv:0808.1335 [hep-th].
  %%CITATION = ARXIV:0808.1335;%%
  
%\cite{Starobinsky:1979ty}
\bibitem{Starobinsky:1979ty}
  A.~A.~Starobinsky,
  %``Spectrum of relict gravitational radiation and the early state of the
  %universe,''
  JETP Lett.\  {\bf 30}, 682 (1979)
  [Pisma Zh.\ Eksp.\ Teor.\ Fiz.\  {\bf 30}, 719 (1979)].
  %%CITATION = ZFPRA,30,719;%%

%\cite{Starobinsky:1980te}
\bibitem{Starobinsky:1980te}
  A.~A.~Starobinsky,
  %``A new type of isotropic cosmological models without singularity,''
  Phys.\ Lett.\  B {\bf 91}, 99 (1980).
  %%CITATION = PHLTA,B91,99;%%

%\cite{Capozziello:2003tk}
\bibitem{Capozziello:2003tk}
  S.~Capozziello, S.~Carloni and A.~Troisi,
  %``Quintessence without scalar fields,''
  Recent Res.\ Dev.\ Astron.\ Astrophys.\  {\bf 1}, 625 (2003)
  [arXiv:astro-ph/0303041].
  %%CITATION = 00638,1,625;%%

%\cite{Carroll:2003wy}
\bibitem{Carroll:2003wy}
  S.~M.~Carroll, V.~Duvvuri, M.~Trodden and M.~S.~Turner,
  %``Is cosmic speed-up due to new gravitational physics?,''
  Phys.\ Rev.\  D {\bf 70}, 043528 (2004)
  [arXiv:astro-ph/0306438].
  %%CITATION = PHRVA,D70,043528;%%
  
   %\cite{Dolgov:2003px}
\bibitem{Dolgov:2003px}
  A.~D.~Dolgov and M.~Kawasaki,
  %``Can modified gravity explain accelerated cosmic expansion?,''
  Phys.\ Lett.\  B {\bf 573}, 1 (2003)
  [arXiv:astro-ph/0307285].
  %%CITATION = PHLTA,B573,1;%%
  
  %\cite{Amendola:2006kh}
\bibitem{Amendola:2006kh}
  L.~Amendola, D.~Polarski and S.~Tsujikawa,
  %``Are f(R) dark energy models cosmologically viable ?,''
  Phys.\ Rev.\ Lett.\  {\bf 98}, 131302 (2007)
  [arXiv:astro-ph/0603703].
  %%CITATION = PRLTA,98,131302;%%
  
  %\cite{Navarro:2006mw}
\bibitem{Navarro:2006mw}
  I.~Navarro and K.~Van Acoleyen,
  %``f(R) actions, cosmic acceleration and local tests of gravity,''
  JCAP {\bf 0702}, 022 (2007)
  [arXiv:gr-qc/0611127].
  %%CITATION = JCAPA,0702,022;%%
  
   %\cite{Chiba:2006jp}
\bibitem{Chiba:2006jp}
  T.~Chiba, T.~L.~Smith and A.~L.~Erickcek,
  %``Solar System constraints to general f(R) gravity,''
  Phys.\ Rev.\  D {\bf 75}, 124014 (2007)
  [arXiv:astro-ph/0611867].
  %%CITATION = PHRVA,D75,124014;%%
  
  %\cite{Amendola:2007nt}
\bibitem{Amendola:2007nt}
  L.~Amendola and S.~Tsujikawa,
  %``Phantom crossing, equation-of-state singularities, and local gravity
  %constraints in $f(R)$ models,''
  Phys.\ Lett.\  B {\bf 660}, 125 (2008)
  [arXiv:0705.0396 [astro-ph]].
  %%CITATION = PHLTA,B660,125;%%
  
  %\cite{Starobinsky:2007hu}
\bibitem{Starobinsky:2007hu}
  A.~A.~Starobinsky,
  %``Disappearing cosmological constant in f(R) gravity,''
  JETP Lett.\  {\bf 86}, 157 (2007)
  [arXiv:0706.2041 [astro-ph]].
  %%CITATION = JTPLA,86,157;%%
 
%\cite{Zhang:2005vt}
\bibitem{Zhang:2005vt}
  P.~Zhang,
  %``Testing $f(R)$ gravity against the large scale structure of the universe,''
  Phys.\ Rev.\  D {\bf 73}, 123504 (2006)
  [arXiv:astro-ph/0511218].
  %%CITATION = PHRVA,D73,123504;%%

    %\cite{Bertschinger:2006aw}
\bibitem{Bertschinger:2006aw}
  E.~Bertschinger,
  %``On the Growth of Perturbations as a Test of Dark Energy,''
  Astrophys.\ J.\  {\bf 648}, 797 (2006)
  [arXiv:astro-ph/0604485].
  
  %\cite{Song:2006ej}
\bibitem{Song:2006ej}
  Y.~S.~Song, W.~Hu and I.~Sawicki,
  %``The large scale structure of f(R) gravity,''
  Phys.\ Rev.\  D {\bf 75}, 044004 (2007)
  [arXiv:astro-ph/0610532].
  %%CITATION = PHRVA,D75,044004;%%

  %\cite{Bertschinger:2008zb}
 \bibitem{Bertschinger:2008zb}
  E.~Bertschinger and P.~Zukin,
  %``Distinguishing Modified Gravity from Dark Energy,''
  arXiv:0801.2431 [astro-ph].
  %%CITATION = ARXIV:0801.2431;%%
  
\bibitem{camb} http://camb.info/

\bibitem{Lewis:1999bs}
  A.~Lewis, A.~Challinor and A.~Lasenby,
  %``Efficient Computation of CMB anisotropies in closed FRW models,''
  Astrophys.\ J.\  {\bf 538}, 473 (2000)
  [arXiv:astro-ph/9911177].
  %%CITATION = ASJOA,538,473;%%

  %\cite{Dunkley:2008ie}
\bibitem{Dunkley:2008ie}
  J.~Dunkley {\it et al.}  [WMAP Collaboration],
  %``Five-Year Wilkinson Microwave Anisotropy Probe (WMAP) Observations:
  %Likelihoods and Parameters from the WMAP data,''
  arXiv:0803.0586 [astro-ph].
  %%CITATION = ARXIV:0803.0586;%%

%\cite{Tegmark:1996bz}
\bibitem{Fisher}
  M.~Tegmark, A.~Taylor and A.~Heavens,
  %``Karhunen-Loeve eigenvalue problems in cosmology: how should we tackle large
  %data sets?,''
  Astrophys.\ J.\  {\bf 480}, 22 (1997)
  [arXiv:astro-ph/9603021].
  %%CITATION = ASJOA,480,22;%%
  
\bibitem{DES}
  http://www.darkenergysurvey.org

\bibitem{LSST}
  http://www.lsst.org

\bibitem{Planck}
http://sci.esa.int/planck

  \bibitem{SNLS} http://www.cfht.hawaii.edu/SNLS/

%\cite{Astier:2005qq}
\bibitem{Astier:2005qq}
  P.~Astier {\it et al.}  [The SNLS Collaboration],
  %``The Supernova Legacy Survey: Measurement of Omega_M, Omega_Lambda and w
  %from the First Year Data Set,''
  Astron.\ Astrophys.\  {\bf 447}, 31 (2006)
  [arXiv:astro-ph/0510447].
  %%CITATION = AAEJA,447,31;%%

\bibitem{NSNF} http://snfactory.lbl.gov/

\bibitem{aldering02}
G.~Aldering \emph{et al}, Survey and Other Telescope Technologies and Discoveries, Proceedings of the SPIE, edited by A.~.J.~Tyson Wolff, Sidney, volume 4836, pp. 61-72 (2002).

 \bibitem{SNAP}
http://snap.lbl.gov

\bibitem{tegmark-pca} %\cite{Tegmark:1997jr}
  A.~J.~S.~Hamilton and M.~Tegmark,
  %``Decorrelating the Power Spectrum of Galaxies,''
  Mon.\ Not.\ Roy.\ Astron.\ Soc.\  {\bf 312}, 285 (2000)
  %[arXiv:astro-ph/9905192].
  %%CITATION = MNRAA,312,285;%%

%\cite{Zhao:2009fn}
\bibitem{Zhao:2009fn}
  G.~B.~Zhao, L.~Pogosian, A.~Silvestri and J.~Zylberberg,
  %``Cosmological tests of GR -- a look at the principals,''
  arXiv:0905.1326 [astro-ph.CO].
  %%CITATION = ARXIV:0905.1326;%%
  
\end{thebibliography}
\end{document}